\documentclass[lettersize,journal]{IEEEtran}
\usepackage{amsmath,amsfonts}
\usepackage[ruled]{algorithm2e}
\usepackage{array}
\usepackage[caption=false,font=normalsize,labelfont=sf,textfont=sf]{subfig}
\usepackage{textcomp}
\usepackage{stfloats}
\usepackage{url}
\usepackage{amssymb}
\usepackage{verbatim}
\usepackage{soul}
\usepackage{graphicx}
\usepackage{cuted}
\usepackage{bm} 
\usepackage{multirow}
\usepackage{cite}

\begin{document}
	\title{From OFDM to AFDM: Enabling Adaptive Integrated Sensing and Communication in High-Mobility Scenarios}
	
	\author{Haoran Yin, \textit{Graduate Student Member, IEEE}, Yanqun Tang, Jun Xiong, \textit{Member, IEEE}, \\ Fan Liu, \textit{Senior Member, IEEE}, Yuanhan Ni, \textit{Member, IEEE}, Qu Luo, \textit{Member, IEEE}, Roberto Bomfin,\\ Marwa Chafii, \textit{Senior Member, IEEE}, Marios Kountouris, \textit{Fellow, IEEE}, Christos Masouros, \textit{Fellow, IEEE}
	\thanks{
		Haoran Yin is with the School of Electronics and Communication Engineering, Sun Yat-sen University, China, and also with the Wireless Research Lab, Engineering Division, New York University (NYU) Abu Dhabi, UAE;
		
		Yanqun Tang (corresponding author) is with the School of Electronics and Communication Engineering, Sun Yat-sen University, China, and also with the Guangdong Provincial Key Laboratory of Sea-Air-Space Communication, China; 
		
		Jun Xiong is with the College of Electronic Science and Technology, National University of Defense Technology, China;
		
		Fan Liu is with the National Mobile Communications Research Laboratory, Southeast University, China;
		
		Yuanhan Ni is with the School of Electronic and Information Engineering, Beihang University, China;
		
		Qu Luo is with the 5G and 6G Innovation centre, Institute for Communication Systems (ICS) of University of Surrey, Guildford, GU2 7XH, UK; 
		
		Roberto Bomfin is with the Wireless Research Lab, Engineering Division, New York University (NYU) Abu Dhabi, UAE;
		
		Marwa Chafii is with the Wireless Research Lab, Engineering Division, New York University (NYU) Abu Dhabi, UAE, and also with NYU WIRELESS, NYU Tandon School of Engineering, New York, USA;
		
		Marios Kountouris is with the Communication Systems Department, EURECOM, France and with the Andalusian Research Institute in Data Science and Computational Intelligence (DaSCI), Department of Computer Science and Artificial Intelligence, University of Granada, Spain;
		
		Christos Masouros is with the Department of Electrical and Electronic Engineering, University College London, UK.
		
	}  	
		
	}
	\markboth{}%
	{Shell \MakeLowercase{\textit{et al.}}: A Sample Article Using IEEEtran.cls for IEEE Journals}
	\maketitle

\begin{abstract}
	\textbf{Integrated sensing and communication (ISAC) is a key feature of next-generation wireless networks, enabling a wide range of emerging applications such as vehicle-to-everything (V2X) and unmanned aerial vehicles (UAVs), which operate in high-mobility scenarios. Notably, the wireless channels within these applications typically exhibit severe delay and Doppler spreads. The latter causes serious communication performance degradation in the Orthogonal Frequency-Division Multiplexing (OFDM) waveform that is widely adopted in current wireless networks. To address this challenge, the recently proposed Doppler-resilient affine frequency division multiplexing (AFDM) waveform, which uses flexible chirp signals as subcarriers, shows great potential for achieving adaptive ISAC in high-mobility scenarios. This article provides a comprehensive overview of AFDM-ISAC. We begin by presenting the fundamentals of AFDM-ISAC, highlighting its inherent frequency-modulated continuous-wave (FMCW)-like characteristics. Then, we explore its ISAC performance limits by analyzing its diversity order, ambiguity function (AF), and Cramér-Rao Bound (CRB). Finally, we present several effective sensing algorithms and opportunities for AFDM-ISAC, with the aim of sparking new ideas in this emerging field.}
\end{abstract}

\section{Introduction}
\label{secI}
\begin{figure}[tbp]
	\centering
    \includegraphics[width=0.450\textwidth,height=0.42\textwidth]{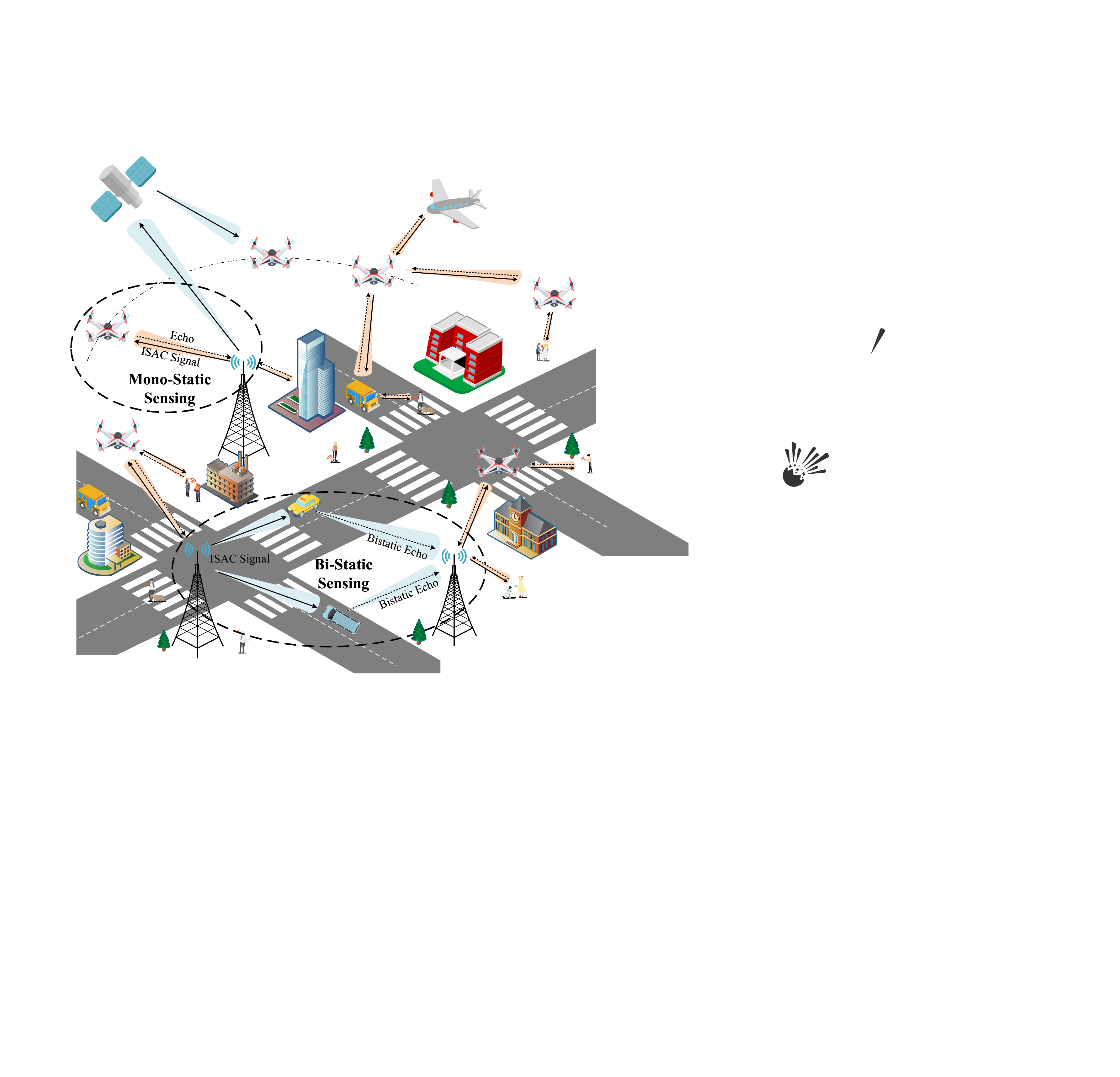}
	\caption{ISAC-enabled ubiquitous connectivity in NGWNs.}
	\label{fig1}
\end{figure}

Next-generation wireless networks (NGWNs) are designed to support cutting-edge applications such as vehicle-to-everything (V2X) communications, low-altitude wireless networks (LAWNs) enabled by unmanned aerial vehicles (UAVs), and satellite networks, as illustrated in Fig.~\ref{fig1}. These transformative applications depend on reliable networking and precise environmental perception in high-mobility scenarios, where wireless channel conditions exhibit rapid variations. Traditionally, communication and sensing have been performed separately using distinct hardware and signal processing frameworks, leading to spectrum contention and mutual interference. In this context, integrated sensing and communication (ISAC) has emerged as a key innovation, enabling the seamless integration of communication and sensing functionalities. This convergence enhances both spectrum and energy efficiency while reducing hardware costs~\cite{bb25.2.4.1}.

Communication-centric ISAC, which leverages communication waveforms to realize sensing functionality, represents a promising approach for achieving effective ISAC due to its inherent compatibility with existing Third Generation Partnership Project (3GPP) standards. In contrast, sensing-centric ISAC, which adapts traditional sensing waveforms to support communication, generally offers lower data rates and higher deployment costs. Owing to the widespread adoption of orthogonal frequency-division multiplexing (OFDM) in modern wireless systems, OFDM-ISAC naturally emerges as an attractive solution for NGWNs. This is supported by the fact that OFDM achieves the lowest average ranging sidelobe under quadrature amplitude modulation (QAM) and phase-shift keying (PSK) constellations~\cite{bb25.7.29.1}. However, in high-mobility scenarios, multipath wireless channels exhibit both significant delay spread and non-negligible Doppler shifts. The latter, induced by the relative motion among the transmitter, scatterers, and receiver, substantially degrades the communication performance of OFDM. These challenges are expected to become even more severe at high-frequency bands used to alleviate spectrum congestion, since Doppler shifts scale linearly with the carrier frequency.

Against this background, numerous new waveforms have been proposed over the past decade to address the aforementioned challenges. One of the most compelling approaches is implementing ISAC in the delay–Doppler (DD) domain by leveraging waveforms such as orthogonal time–frequency space (OTFS), orthogonal delay–Doppler division multiplexing (ODDM), and delay–Doppler alignment modulation (DDAM)~\cite{bb24.03.15.33}. By multiplexing information symbols directly in the DD domain, OTFS and ODDM establish an explicit relationship between the signal–channel interaction and the sensing parameters of targets, thereby offering intuitive insight into ISAC operation.  
Another promising direction involves realizing ISAC through the use of chirp-based signals, which are characterized by a linearly increasing instantaneous frequency over time. This unique time–frequency (TF) spanning property provides chirp signals with high DD sensing resolution, a feature that has long been exploited in radar sensing systems such as frequency-modulated continuous-wave (FMCW) radar. Recently, an innovative chirp-based multicarrier waveform, termed affine frequency-division multiplexing (AFDM), has been proposed for high-mobility communications, attracting growing interest from both academia and industry~\cite{bb24.08.27.2}.

The core concept of AFDM lies in employing a set of chirp signals with a shared and controllable chirp slope as subcarriers by exploiting the \emph{discrete affine Fourier transform} (DAFT), a discrete counterpart of the \emph{affine Fourier transform} (AFT). Notably, the DAFT possesses two inherent tunable chirp parameters that independently control the chirp slope and the instantaneous phase of all chirp subcarriers, respectively, thereby providing AFDM with an unprecedented degree of flexibility~\cite{bb25.01.08.2}.  
In particular, when both chirp parameters are set to zero, the DAFT reduces to the widely adopted \emph{discrete Fourier transform} (DFT), indicating that AFDM is highly backward compatible with OFDM. Moreover, by finely tuning these parameters according to channel conditions, AFDM can fully separate propagation paths with distinct delay or Doppler shifts in the DAFT domain—a property known as \emph{path separability}. This feature not only endows AFDM with optimal diversity but also facilitates efficient channel estimation (CE) and multiuser access in delay–Doppler spread channels. Consequently, with its exceptional flexibility and inherent chirp-based structure, AFDM holds great promise for realizing effective communication-centric ISAC in high-mobility scenarios.

This article provides a comprehensive and intuitive overview of AFDM-ISAC. We begin with the fundamentals of AFDM-ISAC by discussing key performance metrics and signal categories in ISAC systems, followed by a detailed introduction to AFDM-ISAC signal generation. Next, the communication and sensing performance of AFDM-ISAC is evaluated, with emphasis on its achievable diversity order, ambiguity function (AF), and Cramér–Rao bound (CRB). Furthermore, several promising signal processing algorithms for AFDM-ISAC are presented to demonstrate their practical feasibility. Finally, we outline potential future research directions for AFDM-ISAC, aiming to provide valuable insights and inspire further advancements in this emerging field.

\section{Fundamentals of AFDM-ISAC}
In this section, we introduce the fundamentals of AFDM-ISAC, which form the foundation for its performance evaluation and signal processing design in the following sections.

\subsection{Communication and Sensing Performance Metrics} 

\subsubsection{Communication Metrics} 
Reliability and efficiency are two fundamental performance indicators in communication. Reliability measures how accurately communication systems convey information bits from the transmitter to the receiver, which is typically quantified by the bit error ratio (BER) at a given signal-to-noise ratio (SNR). Efficiency, on the other hand, assesses the amount of information that can be precisely conveyed from the transmitter to the receiver under constraints of limited time, frequency, spatial, and energy resources.

\subsubsection{Sensing Metrics}
Resolution and accuracy are two fundamental indicators for evaluating sensing performance. Specifically, sensing resolution reflects the ability to distinguish multiple targets, which is fundamentally determined by the AF of the sensing signal. In contrast, sensing accuracy refers to target detection and parameter estimation, where the first involves identifying the presence of targets, and the second involves estimating the parameters of interest for the detected targets, such as range, velocity, and angle. In particular, target detection accuracy can be quantified using the detection probability and false-alarm probability, while parameter estimation accuracy is typically evaluated using metrics such as the mean squared error (MSE) and the CRB, which will be introduced in detail in the following section.

\subsubsection{ISAC Metrics}
ISAC metrics refer to performance measures that jointly characterize the sensing and communication capabilities of ISAC systems. For example, the ISAC mutual information (MI) quantifies both the information transfer rate over the communication channel and the amount of information that the sensing echoes convey about the targets or the surrounding environment. Furthermore, sensing-related ISAC metrics can be analyzed by accounting for the randomness introduced by the data payload in ISAC signals, such as the expected AF, ergodic MSE, and ergodic CRB, which capture the sensing performance under random ISAC signaling.

\begin{figure*}[htbp]
	\centering
	\includegraphics[width=1\textwidth,height=0.83\textwidth]{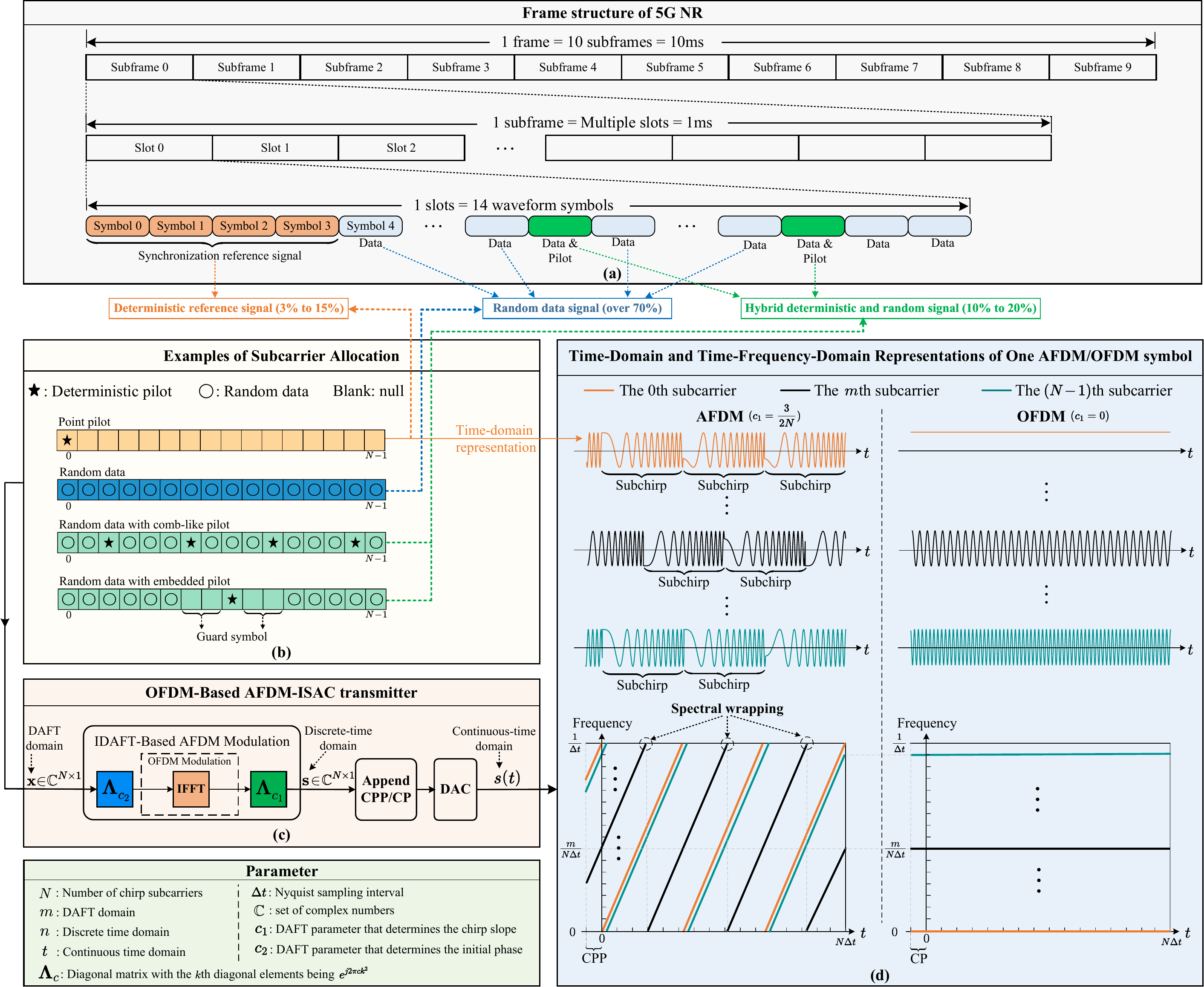}
\caption{(a) Example of the frame structure in 5G NR; (b) examples of subcarrier allocation in AFDM and OFDM; (c) OFDM-based baseband AFDM-ISAC signal generation; and (d) time-domain (real part) and TF-domain representations of a single AFDM/OFDM symbol.}
	\label{fig2}
\end{figure*}

\subsection{Categories of Communication-Centric ISAC Signals}
\label{secII-B}
In a typical 5G New Radio (NR) frame, two core components coexist: (i) a deterministic component, such as preambles and pilots used for multiple access, synchronization, and CE; and (ii) a random component corresponding to the data payload, as illustrated in Fig.~\ref{fig2}(a). Building upon this structure, we heuristically compare three categories of ISAC signals, emphasizing their respective advantages and limitations.

\subsubsection{Deterministic Reference Signal (DRS)}
DRSs refer to waveform symbols composed of reference signals that are fully known to both the transmitter and receiver, such as the synchronization signal block (SSB) for TF synchronization, and the channel state information reference signals (CSI-RS) for channel measurement in 5G NR. These reference signals typically exhibit favorable auto- and cross-correlation properties, enabling accurate and robust mono-static and bi-static sensing, where the transmitter and receiver are co-located in mono-static sensing and spatially separated in bi-static sensing \cite{bb24.9.08.4}, as illustrated in Fig.~\ref{fig1}. However, DRSs generally occupy only a small portion of the 5G NR frame’s TF resources—typically ranging from 3\% to 15\%, depending on channel conditions and numerology configuration—which constrains their ability to achieve high-accuracy, real-time sensing in high-mobility scenarios.

\subsubsection{Random Data Signal (RDS)}
\label{secII-B-2}
RDSs refer to waveform symbols that carry purely random data, such as PSK and QAM symbols, which typically occupy more than 70\% of the TF resources in a 5G NR frame. Despite their information-embedded randomness, these symbols are fully known at the transmitter, thereby offering great potential for real-time mono-static sensing \cite{bb25.01.09.3}. However, the inherent randomness of RDSs inevitably introduces fluctuations in individual sensing realizations, resulting in unstable sensing performance. Consequently, sensing performance should be evaluated in a statistical sense. Notably, although the random data are initially unknown to the bi-static ISAC receiver, bi-static sensing remains feasible provided that the data can be sufficiently recovered.

\subsubsection{Hybrid Deterministic and Random Signal (HDRS)} In addition to DRS and RDS, 5G NR frames also include waveform symbols that embed deterministic pilots within random data symbols, typically arranged in a comb-like subcarrier distribution for CE. Embedding pilots within random data is widely adopted in modulation schemes such as OTFS and AFDM due to their structural simplicity and effectiveness. These signals provide a low-complexity solution for real-time CE, where guard symbols are required to ensure pilot–data isolation \cite{bb24.03.15.33, bb24.08.27.2}. Since these waveform symbols incorporate both deterministic and random components, they are referred to as HDRSs. In 5G NR, HDRSs typically occupy approximately 10\%–20\% of the TF resources, depending on the CE update rate required under varying channel conditions.

\subsection{AFDM-ISAC Signaling}
We now delve into the generation of AFDM-ISAC signals, emphasizing their seamless compatibility with OFDM and close relation to FMCW signals. In AFDM modulation, DAFT-domain data or pilot symbols—denoted by the vector $\mathbf{x} \in \mathbb{C}^{N \times 1}$ and illustrated in Fig.~\ref{fig2}(b)—are multiplexed onto $N$ chirp subcarriers through the inverse DAFT (IDAFT) operation \cite{bb24.08.27.2}. Notably, as depicted in Fig.~\ref{fig2}(c), the IDAFT possesses two intrinsic, tunable chirp parameters and can be efficiently implemented in three successive steps.

Firstly, DAFT-domain chirp precoding is performed, where the input vector $\mathbf{x}$ is multiplied by a diagonal matrix associated with the chirp parameter $c_{2}$. The resulting signal is then processed by an OFDM modulator, i.e., the inverse fast Fourier transform (IFFT) module. Finally, the output of the IFFT is multiplied by another diagonal matrix corresponding to the chirp parameter $c_{1}$, yielding the discrete-time AFDM-ISAC signal $\mathbf{s} \in \mathbb{C}^{N \times 1}$. It is worth noting that the diagonal matrices applied before and after the IFFT module essentially perform element-wise multiplications, ensuring that the computational complexity of the IDAFT remains nearly identical to that of the IFFT in OFDM. Moreover, the OFDM-based structure of the AFDM modulator ensures strong backward compatibility with OFDM, as the IDAFT reduces to the IFFT when both chirp parameters are set to zero—transforming the two diagonal matrices into identity matrices.

In AFDM, the parameter $c_{1}$ is typically configured as an integer multiple of $\frac{1}{2N}$ and governs the chirp slope of all chirp subcarriers, whereas $c_{2}$ is unrestricted and determines their initial phase. This inherent tunability of the chirp parameters in the DAFT domain endows AFDM with remarkable flexibility to tailor its TF distribution for diverse channel conditions, thereby enabling a unified yet customizable air interface for future heterogeneous networks \cite{bb25.01.08.2}. It is also worth noting that when the chirp parameters $(c_{1}, c_{2})$ are both set to $\frac{1}{2N}$, AFDM degenerates into another chirp-based waveform, orthogonal chirp-division multiplexing (OCDM), which can thus be regarded as a special case of AFDM.

After obtaining $\mathbf{s}$, a chirp-periodic prefix (CPP)—which preserves the chirp characteristics of the subcarriers—is appended to the beginning of $\mathbf{s}$ to mitigate the delay spread caused by multipath propagation in wireless channels. The CPP is typically equivalent to the cyclic prefix (CP) used in OFDM systems \cite{bb24.08.27.2}. Subsequently, a digital-to-analog converter (DAC) converts the CPP-appended discrete-time signal into a continuous-time waveform, yielding the AFDM-ISAC signal $s(t)$, whose time-domain and TF-domain representations are shown in Fig.~\ref{fig2}(d).
Notably, all chirp subcarriers exhibit a phenomenon known as spectral wrapping: as their linearly increasing instantaneous frequency reaches the upper limit of the signal bandwidth—defined by the Nyquist pulse-shaping (PS) filter in the DAC—it abruptly wraps around to the lower limit. This effect effectively confines the signal within a practical bandwidth, thereby maintaining high spectral efficiency in AFDM, regardless of the chirp slope configured by the parameter $c_{1}$.
Interestingly, the spectral wrapping behavior implies that each chirp subcarrier consists of multiple subchirps, collectively occupying the entire TF plane of an AFDM-ISAC symbol, as illustrated in Fig.~\ref{fig2}(d). This inherent subchirp periodicity of AFDM subcarriers bears a striking resemblance to FMCW signals, which also comprise multiple chirp components. Therefore, AFDM can be interpreted as a multi-carrier FMCW waveform. As will be discussed later, this FMCW-like property endows AFDM-ISAC signals with several advantages for low-complexity ISAC signal processing.

\section{Performance Evaluation of AFDM-ISAC}
In this section, we comprehensively evaluate the communication and sensing performance limits of AFDM-ISAC in comparison with traditional waveforms, thereby demonstrating its fundamental suitability for ISAC in high-mobility scenarios.

\begin{figure}[tbp]
	\centering
\includegraphics[width=0.470\textwidth,height=0.40\textwidth]{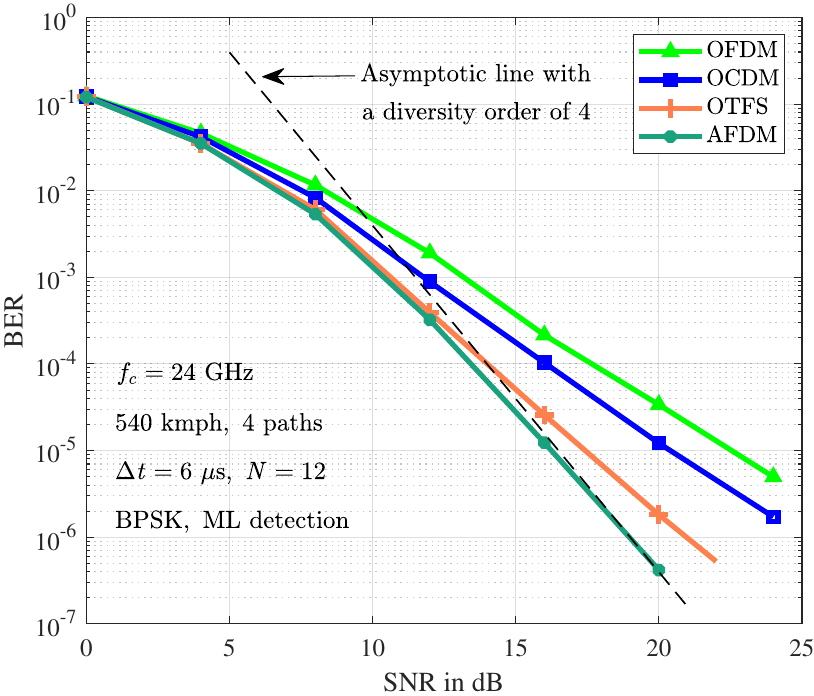}
\caption{Uncoded BER comparison among OFDM, OCDM, OTFS, and AFDM using a ML detector, with carrier frequency $f_{c} = 24$~GHz, $N = 12$, and four propagation paths characterized by a maximum delay of $2\Delta t$ and a maximum Doppler shift of $\frac{1}{N\Delta t}$.}
	\label{fig3}
\end{figure}

\subsection{Achievable Diversity Order}
\label{sec3-A}
We begin by analyzing the BER performance of AFDM-ISAC systems, which can be fundamentally characterized by the achievable diversity order. The diversity order is mathematically defined as the negative slope of the BER–versus–SNR curve on a log–log scale, representing the rate at which the BER decreases as the SNR increases. A higher diversity order corresponds to a steeper BER decline, indicating improved communication reliability. In particular, when the achievable diversity order equals the number of resolvable propagation paths in either the delay or Doppler domain, the system is said to achieve optimal diversity under that channel condition.
Fig.~\ref{fig3} presents the uncoded BER performance of various waveforms under high-mobility conditions, including OFDM, OCDM, OTFS, and AFDM. All waveforms occupy identical TF resources to ensure a fair comparison. The simulation assumes four propagation paths, a maximum speed of 540~km/h, and perfect channel state information at the receiver. BPSK modulation and an optimal ML detector are employed. It can be observed that AFDM achieves the best BER performance among all compared waveforms, as it is the only one attaining optimal diversity, with its BER curve closely following the asymptotic line corresponding to a diversity order of four in the high-SNR region.

\subsection{Ambiguity Function}
We next investigate the sensing resolution performance of AFDM-ISAC signals by analyzing their AF, which is defined as
\begin{equation}
A_{s}(\tau, \nu) \triangleq \int_{-\infty}^{\infty} s(t)s^{*}(t-\tau)e^{-j2\pi\nu t} \mathrm{d}t,
\label{eq24.01.09.6}
\end{equation}
where $s(t)$ denotes the transmitted continuous-time AFDM-ISAC signal, and $\tau$ and $\nu$ represent the delay and Doppler shifts, respectively. Notably, $A_{s}(\tau, \nu)$ quantifies the correlation between $s(t)$ and its time- and frequency-shifted replica characterized by $(\tau, \nu)$ in the DD domain. This function fundamentally determines the output of the matched-filter (MF) operation and serves as a key indicator of the sensing resolution capability of the waveform.

\begin{figure*}[htbp]
	\centering
	\includegraphics[width=1\textwidth,height=0.58\textwidth]{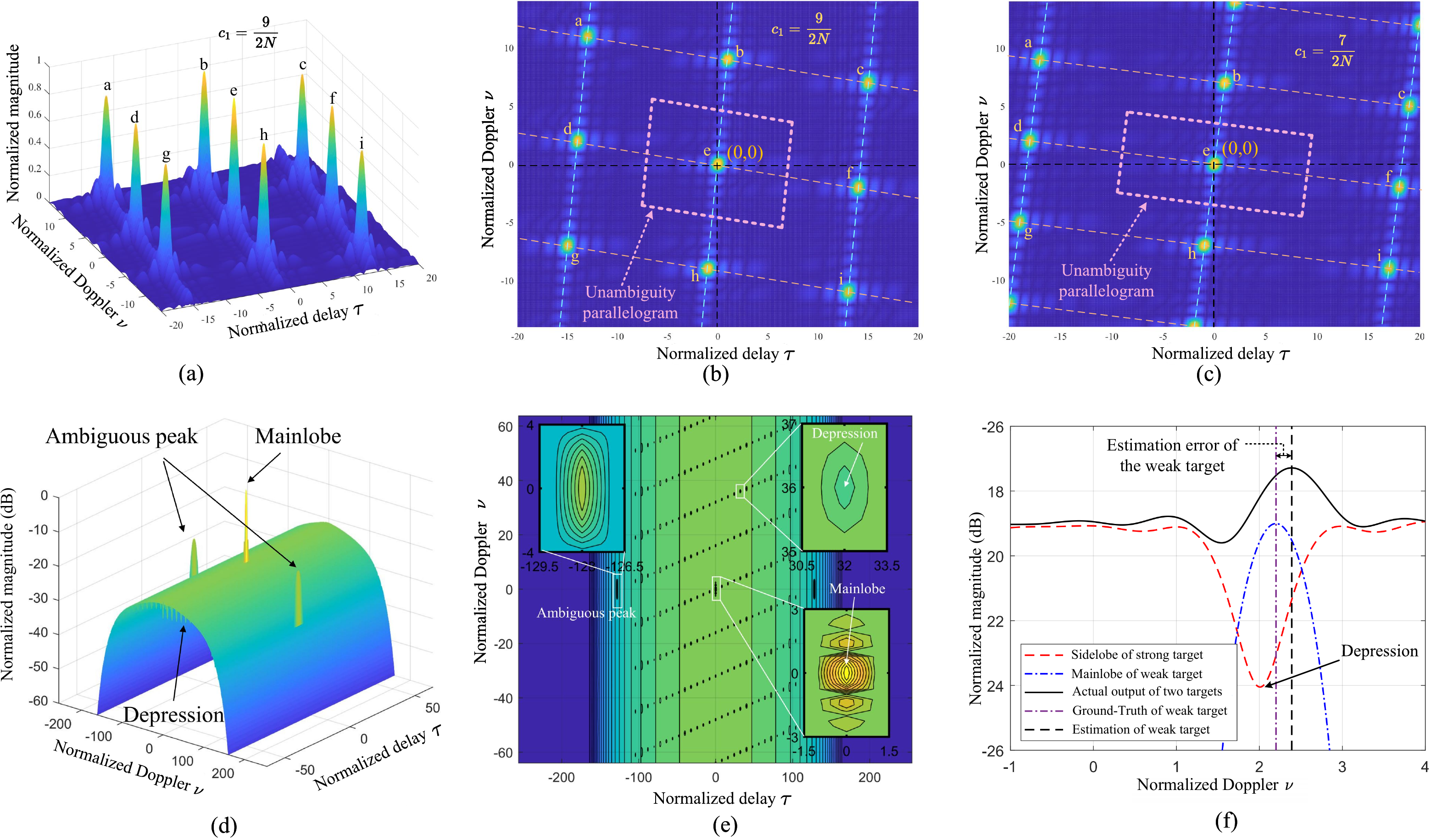}
	\caption{Ambiguity function of AFDM-ISAC signals with $N = 128$: (a) AF of AFDM chirp subcarriers with $c_{1} = \tfrac{9}{2N}$; (b) planform view of Fig.~\ref{fig4}(a); (c) planform view of the AF of AFDM chirp subcarriers with $c_{1} = \tfrac{7}{2N}$; (d) AF of AFDM symbols using 16QAM signaling with $c_{1} = \tfrac{8}{2N}$; (e) planform view of Fig.~\ref{fig4}(d) with contour lines; (f) example illustrating how a strong target can impair the sensing accuracy of a weak target.}
	\label{fig4}
\end{figure*}

\subsubsection{AF of AFDM Chirp Subcarriers}
We first examine the AF of an AFDM chirp subcarrier, which essentially corresponds to sensing with a deterministic point pilot in the DAFT domain, as illustrated in the first case of Fig.~\ref{fig2}(b). Figure~\ref{fig4}(a) depicts the AF of AFDM chirp subcarriers, which exhibits two prominent characteristics: (i) multiple localized, spike-like pulses with low sidelobe levels, and (ii) a periodic distribution of these localized pulses along the rotated Doppler and delay dimensions~\cite{bb24.9.08.105}.

The “spike-like” localized pulses provide highly favorable delay and Doppler sensing resolution, making them particularly advantageous for multi-target detection in high-mobility scenarios. This property arises from the fact that all AFDM chirp subcarriers fully occupy the entire TF resources of the AFDM symbol, as illustrated in Fig.~\ref{fig2}(d). Moreover, the periodic-like pulse distribution results from the subchirp periodicity of AFDM chirp subcarriers, which may introduce target ambiguity in multi-target environments. To address this issue, the authors in~\cite{bb24.9.08.105} proposed an unambiguity parallelogram in the AF of AFDM chirp subcarriers, as shown in Fig.~\ref{fig4}(b), and demonstrated that target ambiguity can be completely avoided when the delay and Doppler shifts of all targets lie within this region. Notably, the area of the unambiguity parallelogram is constant and equal to one, reflecting a fundamental trade-off between the maximum unambiguous range and maximum unambiguous velocity. This trade-off can be flexibly tuned by adjusting the chirp parameter $c_{1}$, which controls the shape of the unambiguity parallelogram, as illustrated in the two cases shown in Fig.~\ref{fig4}(b)–(c).

\subsubsection{AF of AFDM Symbols}
We next examine the AF of AFDM symbols carrying random data, which are expected to constitute the dominant portion of practical AFDM-ISAC transmissions, as discussed in Sec.~\ref{secII-B}. Figures~\ref{fig4}(d)–(e) show the average squared AF of AFDM symbols, obtained by taking the expectation over random data under a practical root-raised-cosine PS filter. It can be observed that, in addition to the thumbtack-like mainlobe and two ambiguous side-lobe peaks, the AF exhibits regular depressions in the sidelobe regions. As highlighted in~\cite{bb24.9.23.3}, the positions of these depressions are determined by the chirp parameter $c_{1}$. Consequently, the presence of strong targets may degrade the sensing accuracy for weaker targets if they fall within these sidelobe depressions, as illustrated in Fig.~\ref{fig4}(f). To mitigate this effect, $c_{1}$ should be adaptively tuned to prevent weak targets from aligning with the sidelobe depressions caused by strong targets, thereby enhancing both detection and parameter estimation performance.

\subsection{Cramér–Rao Bound}
\begin{figure}[tbp]
	\centering
    \includegraphics[width=0.470\textwidth,height=0.7\textwidth]{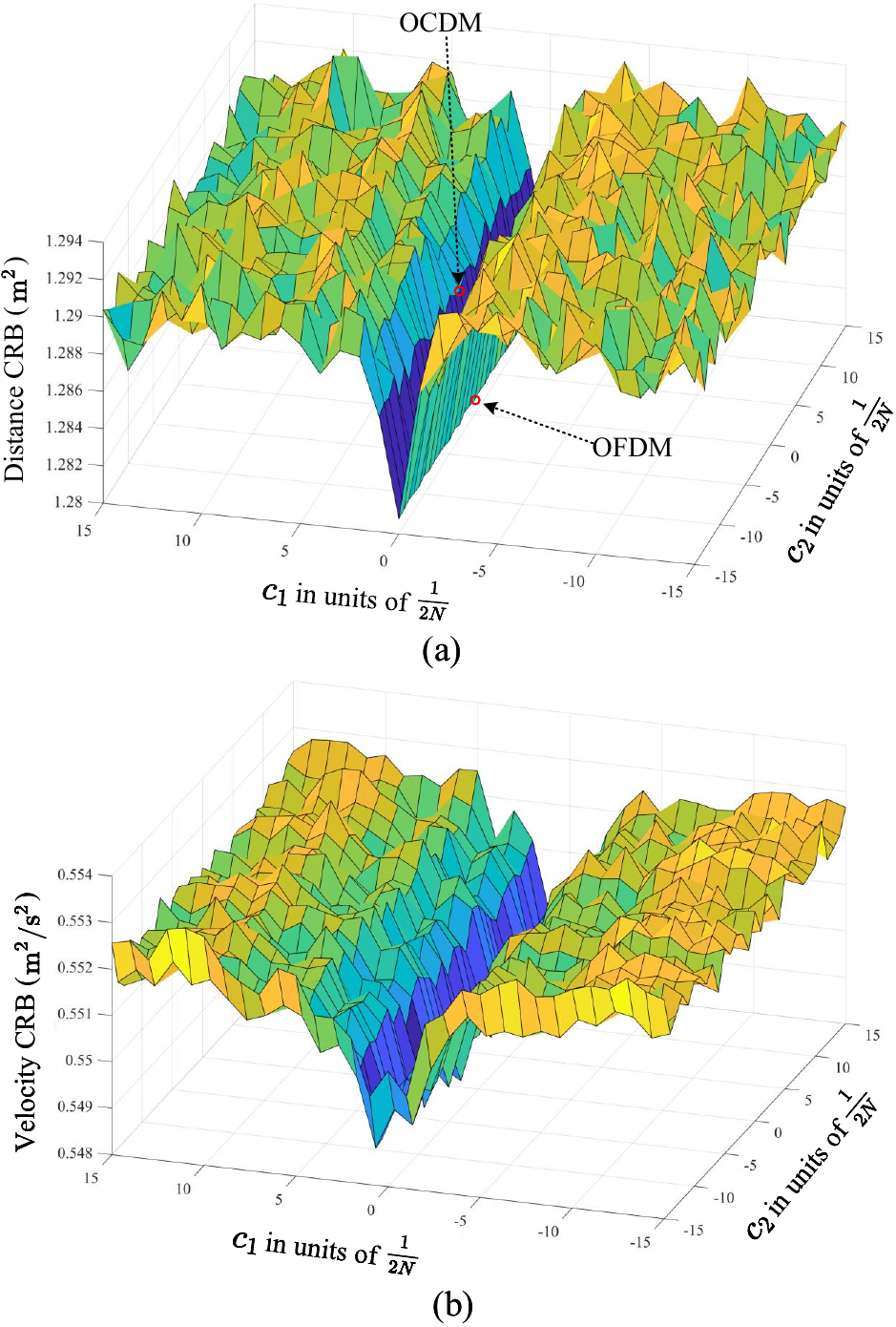}
\caption{Averaged CRBs of AFDM-ISAC symbols with random data under different chirp parameter configurations. A single target with a round-trip distance of 1~km and a velocity of 300~km/h is considered. The carrier frequency is 60~GHz, modulation is 4QAM, $N = 128$, $\Delta t = 0.78~\mu$s, and the SNR is 10~dB. (a) Averaged distance CRB; (b) averaged velocity CRB.}
	\label{fig5}
\end{figure}

The CRB defines the theoretical lower limit on the variance of any unbiased estimator of parameters—such as the delay and Doppler shifts of targets at a given SNR—and serves as a key benchmark for assessing estimation accuracy~\cite{bb25.07.23.2}. Figure~\ref{fig5} illustrates the distance and velocity CRBs of AFDM-ISAC symbols, averaged over random data for various chirp parameter configurations, obtained through 10,000 Monte Carlo simulations. The distance CRBs of OFDM and OCDM are also included for comparison.
It can be observed that both distance and velocity CRB curves exhibit slight fluctuations across $(c_{1}, c_{2})$, with $c_{1}$ exerting a more pronounced influence than $c_{2}$. OFDM achieves the lowest distance CRB; however, the distance CRB of AFDM with $(c_{1}, c_{2}) \neq (0, 0)$ is only marginally higher, by less than 0.5\%. Furthermore, the magnitude of fluctuation in both distance and velocity CRBs across $(c_{1}, c_{2})$ is negligible (less than 0.2\% of their average values). These results indicate that AFDM-ISAC provides robust and stable sensing accuracy across all chirp parameter configurations.

\section{Signal Processing for AFDM-ISAC}
In this section, we introduce several promising signal processing algorithms developed for AFDM-ISAC, emphasizing their unique advantages, implementation feasibility, and potential to enhance integrated sensing and communication performance.

\begin{figure*}[htbp]
\centering
\includegraphics[width=1\textwidth,height=1.29\textwidth]{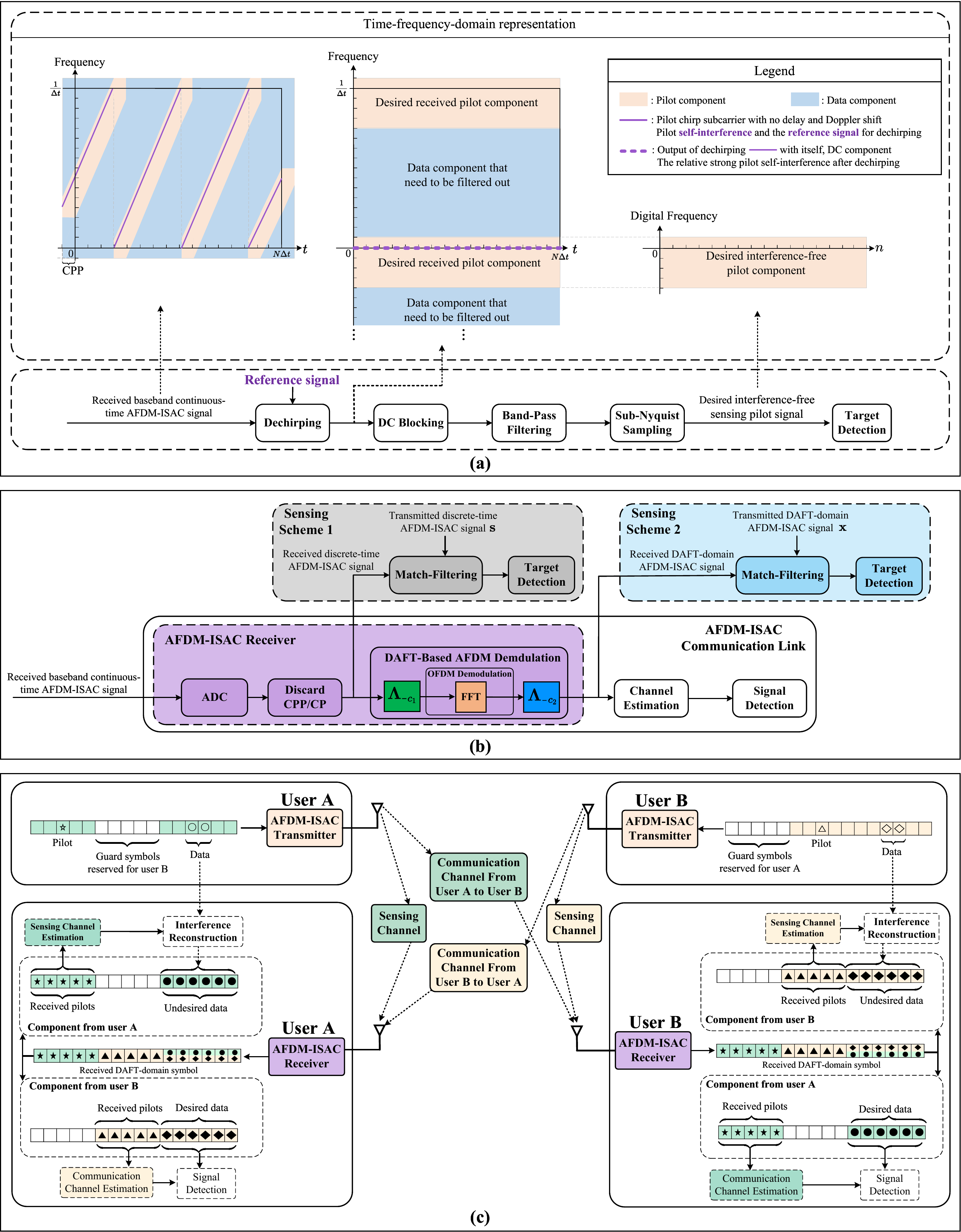}
\caption{Block diagrams of three signal processing algorithms for AFDM-ISAC: (a) dechirping-aided sensing; (b) matched-filtering-based sensing; and (c) DAFT-domain full-duplex ISAC, where the AFDM-ISAC transmitter and receiver modules are illustrated in Fig.~\ref{fig2}(c) and Fig.~\ref{fig6}(b), respectively.}
\label{fig6}
\end{figure*}

\subsection{Dechirping-Aided Sensing}
One of the major challenges in monostatic sensing lies in suppressing the self-interference (SI) at the sensing receiver, which arises from power leakage between the co-located transmitter and receiver and is typically much stronger than the desired echo signals. In AFDM-ISAC, efficient SI cancellation can be realized by leveraging the FMCW-like characteristics of AFDM chirp subcarriers, as illustrated in Fig.~\ref{fig6}(a).

Consider that an embedded pilot with guard symbols is employed in the transmitted AFDM-ISAC symbol, as illustrated at the bottom of Fig.~\ref{fig2}(b). The received continuous-time AFDM-ISAC signal consists of two components: a pilot component for sensing and a data component for communication. Owing to the insertion of guard symbols, these components can be perfectly separated in the time-frequency domain, as indicated in orange and blue, respectively. To extract the pilot component, both the data component and the strong pilot self-interference (SI)—represented by the purple solid line—must be removed.  
This can be achieved by first dechirping the received AFDM-ISAC signal using the SI chirp subcarrier, accomplished by multiplying the received signal with the conjugate of the transmitted SI chirp subcarrier~\cite{bb24.03.15.5}. After dechirping, the pilot and data components become separable in the frequency domain, while the pilot SI is converted into a direct-current (DC) component. The pilot SI can then be effectively eliminated using a DC blocker, and the data component can be filtered out directly. The desired interference-free pilot signal is subsequently obtained by performing \emph{sub-Nyquist sampling} on the filtered output at a rate significantly lower than the bandwidth of the received ISAC signal.  
This dechirping-aided AFDM-ISAC scheme achieves extremely low-cost and low-complexity SI cancellation, offering a substantial advantage over conventional ISAC systems based on OFDM, OCDM, or OTFS waveforms.

\subsection{Matched-Filtering-Based Sensing}
\label{sec4-B}
Matched filtering refers to calculating the correlation between the received ISAC signal and the transmitted ISAC signal with delay-Doppler shifts, which effectively measures the energy of the latter within the former and is fundamentally determined by the AF of the transmitted ISAC signal. In monostatic sensing, MF in AFDM-ISAC can be performed either in the discrete-time domain or in the DAFT domain, as illustrated by Scheme~1 and Scheme~2 in Fig.~\ref{fig6}(b)~\cite{bb2022.11.11.2}. This demonstrates that the MF-based sensing link in AFDM-ISAC can be seamlessly integrated with the communication link by reusing existing modules such as the analog-to-digital converter (ADC) and AFDM demodulator. Moreover, the DAFT-based AFDM demodulation can be efficiently implemented by leveraging the fast Fourier transform (FFT)-based OFDM demodulator. Therefore, the MF-based sensing approach offers high hardware efficiency and implementation simplicity, enhancing the overall integration of communication and sensing functionalities.

\subsection{DAFT-Domain Full-Duplex ISAC}
Full-duplex ISAC enables the simultaneous transmission and reception of ISAC signals over the same frequency band, potentially doubling the spectral efficiency compared with conventional half-duplex systems, as illustrated in Fig.~\ref{fig6}(c). Consider two users communicating bidirectionally while performing monostatic sensing on the same frequency band. In this scenario, user~A's ISAC receiver receives two types of signals: (i) the ISAC echo originating from user~A's own transmitter, propagating through the A-to-A sensing channel; and (ii) the communication signal transmitted by user~B, propagating through the B-to-A communication channel. These two signal components interfere with each other and must be effectively separated. This separation can be accomplished in the DAFT domain by exploiting AFDM’s inherent path separability~\cite{bb25.07.23.1}.

As illustrated at the top of Fig.~\ref{fig6}(c), in addition to employing an embedded pilot as shown in Fig.~\ref{fig2}(b), a guard band should be allocated within user~A's transmitted AFDM-ISAC symbol to enable pilot separation for user~B, and vice versa. Consequently, the received pilots originating from the local user and the remote user become separable in the DAFT domain. The former is utilized for monostatic sensing, where the undesired data component from the local user can be reconstructed and subsequently subtracted, while the latter is exploited to estimate the communication channel, thereby enabling interference-free data detection from the remote user. As a result, full-duplex ISAC can be efficiently realized in AFDM, offering significantly enhanced spectral efficiency.

\section{Challenges and Opportunities}
In this section, we highlight several key challenges and potential opportunities in AFDM-ISAC that warrant further investigation and development.

\subsection{Scenario-Adaptive ISAC Frame Design}
As discussed in Sec.~\ref{secI}, NGWNs must support efficient ISAC across a wide range of application scenarios, each with distinct requirements for communication—such as BER and spectral efficiency—and for sensing—such as resolution and accuracy. A promising approach to achieve this objective is the adoption of scenario-adaptive ISAC frames, which dynamically coordinate the allocation of DRS, RDS, and HDRS components and the constellation design according to real-time channel conditions. In particular, the two tunable chirp parameters in AFDM-ISAC offer unprecedented symbol-level adaptability, enabling real-time reconfiguration to accommodate rapid channel variations in high-mobility environments.

\subsection{MIMO-AFDM-ISAC}
The multiple-input multiple-output (MIMO) technique significantly enhances the multiplexing and diversity gains of communication systems while enabling precise beamforming for accurate azimuth and elevation angle estimation—both of which are essential for fully determining target locations. Therefore, a comprehensive performance analysis of MIMO-AFDM-ISAC, encompassing delay–Doppler–angle-domain diversity, AF, and CRB, is of great importance. Moreover, the design of low-complexity, low-latency signal processing algorithms that approach the theoretical ISAC performance limits is crucial for effectively implementing MIMO-AFDM-ISAC in high-mobility scenarios with stringent real-time processing requirements~\cite{bb24.9.08.104}.

\subsection{High-Frequency ISAC}
To support higher data rates for data-intensive communications and achieve finer ranging resolution for precise target localization in NGWNs, larger bandwidths are required. Consequently, ISAC systems must operate in higher frequency bands to alleviate spectrum congestion, such as the millimeter-wave (mmWave) and terahertz (THz) ranges. The dechirping-based sensing mechanism of AFDM-ISAC requires only sub-Nyquist sampling, thereby substantially reducing the ADC burden and making it particularly suitable for high-frequency ISAC. Furthermore, the inherent optimal diversity of AFDM-ISAC endows it with strong resilience and robustness against impairments such as severe path loss, Doppler spreads, carrier frequency offset (CFO), and phase noise (PN) that are prominent in high-frequency bands.

\subsection{ISAC Security}
The growing volume of data exchanged in NGWNs, including users’ personal information and control commands across heterogeneous devices, raises critical security concerns. Moreover, sensitive information such as user location and behavioral patterns is vulnerable to interception by unauthorized ISAC receivers that may exploit the deterministic components of transmitted ISAC signals. In this context, the FMCW-like, TF-spanning property of AFDM chirp subcarriers provides superior resistance to interference and eavesdropping compared with the sinusoidal subcarriers used in OFDM. In addition, as revealed in \cite{bb25.01.08.2}, AFDM-ISAC inherently exhibits strong security due to the high flexibility in chirp parameter selection, which can serve as cryptographic keys for ISAC signal encryption. Nevertheless, fully exploring and exploiting the potential of the AFDM waveform for ISAC security remains an important open research challenge.

\subsection{ISAC Channel Sounding}
Unlike radar sensing, which aims to determine the instantaneous state information of targets, channel sounding focuses on measuring the statistical characteristics of wireless channels, such as the power delay profile and Doppler power spectrum of multipath components, to enable accurate channel modeling. In high-mobility scenarios, frequent channel sounding is essential to track dynamic channel variations, which can consume substantial time–frequency–energy resources \cite{bb24.9.08.103}. A promising approach to mitigate this challenge is to reuse the pilots embedded within the ISAC frame. Compared with OFDM pilot subcarriers, the FMCW-like, tunable AFDM pilot chirp subcarriers provide superior DD sensing resolution, thereby enabling real-time and accurate channel sounding without introducing additional overhead.

\subsection{Prototype and Testbeds}
Beyond theoretical analysis and signal processing design, developing practical prototypes and testbeds for AFDM-ISAC systems is essential to facilitate real-world implementation. Compared with the transmitter, the AFDM-ISAC receiver poses greater challenges, as it must perform TF synchronization, CE, and signal detection, in addition to accurate target detection and parameter estimation to ensure robust ISAC performance. In particular, computational complexity and processing latency warrant priority consideration when selecting signal processing schemes, as these factors directly impact the feasibility of real-time operation. Moreover, further research is needed to enhance the integration of AFDM-ISAC prototypes with existing OFDM platforms, since such integration largely determines the cost and scalability of large-scale deployment.

\section{Conclusion}
AFDM-ISAC represents a promising solution for NGWNs to realize efficient ISAC in high-mobility scenarios. This advantage stems from its inherent adaptability, robust sensing capability, optimal delay-Doppler-resilient communication performance, and strong backward compatibility with OFDM.
In this article, we presented a comprehensive overview of AFDM-ISAC. We first introduced the OFDM-based AFDM-ISAC signal generation process, emphasizing its tunable flexibility and the FMCW-like characteristics of its chirp subcarriers. Then, we theoretically evaluated its performance in terms of optimal diversity order, “spike-like” and adjustable ambiguity function, and stable Cramér–Rao bound. Furthermore, we demonstrated that AFDM-ISAC offers several key advantages, including low-cost self-interference cancellation, high hardware reusability, and support for high-spectral-efficiency full-duplex ISAC. Finally, we outlined several promising research directions to guide the future development of AFDM-ISAC systems.

\vfill
\end{document}